\documentclass[10pt,superscriptaddress,twocolumn,amsmath,amssymb,aps,prl,showpacs]{revtex4}
\usepackage{mathrsfs}
\usepackage{graphicx}
\usepackage{dcolumn}
\usepackage{bm}
\usepackage{amssymb}
\usepackage{amsmath}
\usepackage{paralist}
\usepackage{color}
\DeclareMathOperator\Tr{Tr}
\DeclareMathOperator\MHz{MHz}

\begin{document}

\title{Exact Entanglement dynamics in Three Interacting Qubits}
\author{WenBin He}
\affiliation{State Key Laboratory of Magnetic Resonance and Atomic and Molecular Physics,
Wuhan Institute of Physics and Mathematics, Chinese Academy of Sciences, Wuhan 430071, China}
\affiliation{University of Chinese Academy of Sciences, Beijing 100049, China}

\author{Xi-Wen Guan~\footnote{Correspondence author. Email: xiwen.guan@anu.edu.au}
%  Emai: xiwen.guan@anu.edu.au; hewenbing1991@163.com;
}
\affiliation{State Key Laboratory of Magnetic Resonance and Atomic and Molecular Physics,
Wuhan Institute of Physics and Mathematics, Chinese Academy of Sciences, Wuhan 430071, China}
\affiliation{Center for Cold Atom Physics, Chinese Academy of Sciences, Wuhan 430071, China}
\affiliation{Department of Theoretical Physics, Research School of Physics and Engineering,
Australian National University, Canberra ACT 0200, Australia}

\date{\today}


\begin{abstract}
Motivated by recent experimental study on  coherent dynamics transfer in three interacting atoms or electron spins \cite{Barredo:2015,Rosenfeld:2018}, here we  study entanglement entropy transfer  in three interacting qubits.
We analytically calculate time evolutions of wave function,  density matrix and entanglement of the system. 
 We find that   initially entangled two qubits may alternatively  transfer their entanglement entropy  to other two qubit pairs.
 So that  dynamical evolution of  three interacting qubits  may produce a genuine three-partite entangled state through  entanglement entropy transfers. 
In particular, different pairwise interactions of the three qubits endow symmetric and asymmetric evolutions of the entanglement transfer,  characterized by the quantum mutual information and concurence. 
 Finally, we discuss  an experimental proposal of three Rydberg atoms  for testing  the entanglement dynamics transfer of this kind. 
\end{abstract}
\maketitle

%%%%%%%%%%%%%%%%%%%%%%%%%%%%%%%%%%%%%%%%%

Entanglement is a  fundamental  but rather mysterious phenomenon in quantum many-body physics.  
It has become an essential  theme  in the study of update quantum metrology. 
Due to recent developments of  experimental technology,  some  entangled states of spins, electrons and atoms  can be created  in laboratory.
Such entangled states become  important resources for  high precision measurements  in quantum information and quantum metrology \cite{APolkRMP, Fazio,Qmetro}. 
Very recently,  many experimental works on controlling few qubits were reported, by using Rydberg atom \cite{DJwang,Barredo:2015,MSZhan}, superconduct circuit  \cite{nature},  quantum dot \cite{GcGuo} and a single nitrogen vacancy (NV) center electrons \cite{Rosenfeld:2018}.  
However, quantum entanglement still imposes a big theoretical and experimental challenge. 
From a theoretical point of view, one still does not know how to properly characterise three body entanglement.
In experiment, it is very  difficult to create high quality entangled states of multiple particles due to decoherence, noise and environment  fluctuations etc.
In this scenario, the study of coherent dynamics transfer among entangled qubits, spin diffusion in bath and entanglement entropy become  an important theme of physical interest. 

In this short communication, we present exact entanglement dynamics of three interacting qubits. 
We find that the entanglement entropy transfer and  the genuine three-partite entanglement state  can be  generated  in  dynamical evolution of three qubits with pairwise interaction.
Different pairwise interactions in  the three qubits endow symmetric and asymmetric evolutions of entanglement entropy and concurrence, see  Fig.\ref{fig2:dynamics} and Fig.\ref{fig4:homo}.
Finally, we also discuss an experimental proposal of three Rydberg atoms to test such a kind of  entanglement dynamics transfers. 

\begin{figure}[h]
\begin{center}
\includegraphics[scale=0.4]{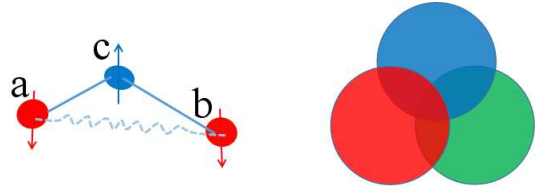}
\end{center} 
\caption{Left panel: Schematic  diagram of three interacting qubits, solid lines represent the interaction between the blue spin $c$ and red spins $a$ or $b$. The  curved line represents the  entanglement between $a$ and $b$. Right panel: Schematic diagram of the  entangled state manifold for the three interacting  qubits. The mutual entanglements are symbolized by three color regions. The non-intersecting regions represent the separable state, the intersecting regions represent the entangled states.}
\label{fig1:schematic}
\end{figure}

{\em Entanglement measures.} Without loosing generality, here we consider  the dynamical evolution of three interacting qubits by choosing an entangled qubit pair  $ab$ as the  initial state
\begin{equation}
\vert \Phi_{0} \rangle=\frac{1}{\sqrt{2}}(\vert \uparrow \downarrow \rangle_{ab}+\vert  \downarrow  \uparrow\rangle_{ab}) \otimes \vert \uparrow\rangle_{c}.
\label{Istate}
\end{equation}
For our convenience, we write initial state in the following  form
\begin{equation}
\vert \Phi_{0} \rangle=\frac{1}{\sqrt{2}}(\mathbf{s_{a}^{-}}+\mathbf{ s_{b}^{-}}) \vert \Uparrow\rangle=\frac{1}{\sqrt{2}}(\vert a \rangle +\vert b \rangle )
\end{equation}
with a notation  $\vert x \rangle=s_{x}^{-}\vert \Uparrow \rangle$, where $s_{x}^{-}$ is spin-$1/2$ lowering operator and $\vert \Uparrow \rangle= \vert \uparrow\uparrow\uparrow \rangle$. See  Fig.~\ref{fig1:schematic} left panel: two entangled qubits (red spins) have interaction with the  third qubit (the blue one).
Under a unitary time evolution,   the wave function at arbitrary time can be written as 
\begin{equation}
\vert \psi(t) \rangle=e^{-iHt} \vert \Phi_{0}\rangle.
\end{equation}
We can also  derive the density matrix of the model  from the above wave function
\begin{equation}
\rho_{s}= \vert \psi(t) \rangle \langle \psi(t) \vert.
\end{equation}
The density matrix  is the key quantity  to signal the  entanglement dynamics transfer. 
In order to achieve this end, we calculate the  quantum mutual information and the concurrence \cite{Nielsen, Wootters97,Wootters98, Wootters00}.
For example, the quantum mutual information and the concurrence of two  qubits  $a$ and $b$  are  defined by 
\begin{eqnarray}
&S(a:b)=S_{a}+S_{b}-S(a,b), \label{Dfsm} \\
&\mathcal{C}_{ab}=\max(0,\lambda_{1}-\lambda_{2}-\lambda_{3}-\lambda_{4}), \label{DfC}
\end{eqnarray}
respectively. 
In the above equation,
the Von Neumann entropy  is given by $S(a,b)=-\Tr[\rho_{ab} \log_{2}\rho_{ab}]$, and $\{\lambda_{j} \}$ are square roots of the eigenvalues of  the non-Hermitian matrix $\rho_{ab} \tilde{\rho}_{ab}$ in decreasing order, here $\tilde{\rho}_{ab}$ is defined by 
\begin{equation*}
\tilde{\rho}_{ab}=(\sigma_{y} \otimes \sigma_{y})\rho^{*}_{ab}(\sigma_{y} \otimes \sigma_{y}).
\end{equation*}
While $\rho_{x}$ denotes the reduced density matrix of a single qubit $x$. 

Although  forementioned entanglement measures are defined for two qubits, we can use the entanglement entropies  of three qubit pairs to witness the three-qubit entanglement, see  Fig.~\ref{fig1:schematic} right panel, in which  three colour regions to symbolize  the state manifold of three qubits. 
We observe  the coexistence of the entanglement entropies of  three qubit pairs $ab$, $ac$ and $bc$. 
The coexistence region  presents a three-partite entanglement state, see Fig.\ref{fig2:dynamics}.
We shall quantitatively study such mutual entanglement entropies below. 

{\em Inhomogeneous interacting qubits.} 
Let's first consider the  entanglement dynamics of three interacting qubits with different spin exchange coupling, described by the Hamiltonian 
\begin{equation}
H=2[A_{a}\mathbf{s_{a}} \mathbf{s_{c}}+A_{b} \mathbf{s_{b}} \mathbf{s_{c}}]. \label{Ham}
\end{equation}
Here  $A_{a,b}$ denote the spin exchange strengths for the  qubit pairs $ac$ and and $bc$, respectively. 
The Hamiltonian (\ref{Ham})  closely relates to the central spin model \cite{J.Dukelsky2004, H.-Q. Zhou, Dloss02, Dloss03, Bortz07,AFprb} with the particle number $N=3$ and  magnetic field $B=0$.  Here the qubit $c$ play the role of the central spin. 
For our convenience,  we introduce parameters $A_{j}=1/(\epsilon_{c}-\epsilon_{j})$, here $\epsilon_{c}=0$. We use the  Bethe ansatz eigenfunction of the central spin model to investigate the time evolution of the system, namely, 
\begin{equation}
\vert \lbrace \nu \rbrace  \rangle=\prod_{\alpha=1}^{M}B_{\nu_{\alpha}} \vert \Uparrow \rangle=\prod_{\alpha=1}^{M}\sum_{j= a,b,c} \frac{s_{j}^{-}}{\nu_{\alpha}-\epsilon_{j}}  \vert \Uparrow \rangle.
\end{equation}
Here $ j= a,b,c$ stands for the bath spins $ a,b$, central spin $c$ and  $M$ is the number of  down-spins. 
The  spectrum parameters $\lbrace \nu\rbrace $ satisfy the Bethe ansatz equations
\begin{equation}
 \sum_{ j= a,b,c} \frac{1}{\nu_{\alpha}-\epsilon_{j}}=\sum_{\beta \neq \alpha , \beta=1}^{M} \frac{2}{\nu_{\alpha}-\nu_{\beta}},
 \label{BAE}
\end{equation}
with $\alpha =1,\ldots, M$. 
 The Bethe  ansatz equations (\ref{BAE}) have $C_{N}^{M}$ sets of solutions in the Hilbert subspace. 
 For our case $M=1$, the eigenergy of the three qubits system reads  
 \begin{equation}
    E=\frac{1}{2} \sum_{j=a,b} \frac{1}{\epsilon_{c}-\epsilon_{j}}-\frac{1}{\epsilon_{c}-\nu},
 \end{equation} 
 where the  Bethe ansatz parameter $\nu$ satisfy the following equation
\begin{equation}
\frac{1}{\nu-\epsilon_{a}}+\frac{1}{\nu-\epsilon_{b}}+\frac{1}{\nu}=0,
\end{equation}
that gives the solutions 
\begin{align*}
&\nu_{1,2}=\frac{1}{3} \left[ (\epsilon_{a}+\epsilon_{b}) \pm \sqrt{\epsilon_{a}^{2}+\epsilon_{b}^{2}-\epsilon_{a}\epsilon_{b}} \right] , \\
&\nu_{3}=\infty. 
\end{align*}
The solutions look rather simple, but  indeed encode a rich quantum dynamics of three interacting qubits. 

The initial state Eq. (\ref{Istate}) belongs to the subspace with $M=1$. 
With the help of the above solutions, the wave function at arbitrary time can be obtained through the unitary evolution 
\begin{equation}
\vert \psi(t) \rangle=e^{-iHt} \vert \Phi_{0}\rangle=\sum_{k} \vert \phi_{k} \rangle \langle \phi_{k} \vert \Phi_{0} \rangle  e^{-iE_{k}t}.
\end{equation}
Here orthonormalized eigenfunction $\vert \phi_{k} \rangle=N_{\nu_{k}} \vert \nu_{k}\rangle$, where $N_{\nu_{k}}$ is normalization factor  
\begin{equation*}
\frac{1}{{\vert N_{\nu_{k}} \vert} ^2} \delta_{k,k^{\prime}}=\sum_{j=a,b,c}\frac{1}{(\nu_{k}-\epsilon_{j})(\nu_{k^{\prime}}-\epsilon_{j})}.
\end{equation*}
By a straightforward calculation of the overlap between eigenfunction and initial state, we obtain the time evolution of the wave function 
\begin{equation}
\vert \psi(t) \rangle=\frac{1}{\sqrt{2}}\sum_{j= a,b,c} C_{j}(t)\vert j \rangle, \label{wave-func}
\end{equation}
where coefficiences $C_{j}$ read
\begin{equation*}
C_{j}(t)=\sum_{k} \vert N_{\nu_{k}}\vert^2 \left[ \frac{1}{\nu_{k}-\epsilon_{a}}+\frac{1}{\nu_{k}-\epsilon_{b}} \right] \frac{e^{-iE_{k}t}}{\nu_{k}-\epsilon_{j}}.
\end{equation*}

The next key step is to calculate the density matrix of system $\rho_{s}= \vert \psi(t) \rangle \langle \psi(t) \vert $. 
Again, using the wave function (\ref{wave-func}),   we can obtain the density matrix 
\begin{equation}
 \rho_{s}= 
\left( \begin{array}{clr}
{A(t)}/{2} & {D(t)}/{2} & {E(t)}/{2}\\
{D(t)^{*} }/{2}& {B(t)}/{2} & {F(t)}/{2}\\
{E(t)^{*} }/{2}& {F(t)^{*} }/{2}&{C(t)}/{2} 
\end{array}  \right), 
\end{equation}
where the  six coefficients are given by 

\begin{widetext}
\begin{align*}
A(t)=&\sum\limits_{k} \sum\limits_{k^{\prime}}{\vert N_{\nu_{k}} \vert} ^2 {\vert N_{\nu_{k^{\prime}}} \vert} ^2 \left[  \sum\limits_{\alpha=a,b}\frac{1}{\nu_{k}-\epsilon_{\alpha}}\right]    \left[ \sum\limits_{\alpha=a,b}\frac{1}{\nu_{k^{\prime}}-\epsilon_{\alpha}} \right] \frac{1}{(\nu_{k}-\epsilon_{a})(\nu_{k^{\prime}}-\epsilon_{a})}\cos(\omega_{k k^{\prime}}t), \\
B(t)=&\sum\limits_{k} \sum\limits_{k^{\prime}}{\vert N_{\nu_{k}} \vert} ^2 {\vert N_{\nu_{k^{\prime}}} \vert} ^2 \left[ \sum\limits_{\alpha=a,b}\frac{1}{\nu_{k}-\epsilon_{\alpha}} \right]     \left[\sum\limits_{\alpha=a,b}\frac{1}{\nu_{k^{\prime}}-\epsilon_{\alpha}} \right]  \frac{1}{(\nu_{k}-\epsilon_{b})(\nu_{k^{\prime}}-\epsilon_{b})}\cos(\omega_{k k^{\prime}}t), \\
C(t)=&\sum\limits_{k} \sum\limits_{k^{\prime}}{\vert N_{\nu_{k}} \vert} ^2 {\vert N_{\nu_{k^{\prime}}} \vert} ^2  \left[ \sum\limits_{\alpha=a,b}\frac{1}{\nu_{k}-\epsilon_{\alpha}}\right] \left[\sum\limits_{\alpha=a,b}\frac{1}{\nu_{k^{\prime}}-\epsilon_{\alpha}}  \right]\frac{1}{(\nu_{k}-\epsilon_{c})(\nu_{k^{\prime}}-\epsilon_{c})}\cos(\omega_{k k^{\prime}}t), \\
D(t)=&\sum\limits_{k} \sum\limits_{k^{\prime}}{\vert N_{\nu_{k}} \vert} ^2 {\vert N_{\nu_{k^{\prime}}} \vert} ^2 \left[ \sum\limits_{\alpha=a,b}\frac{1}{\nu_{k}-\epsilon_{\alpha}}\right] \left[\sum\limits_{\alpha=a,b}\frac{1}{\nu_{k^{\prime}}-\epsilon_{\alpha}} \right] \frac{1}{(\nu_{k}-\epsilon_{a})(\nu_{k^{\prime}}-\epsilon_{b})}e^{ i\omega_{k k^{\prime}}t}, \\
E(t)=&\sum\limits_{k} \sum\limits_{k^{\prime}}{\vert N_{\nu_{k}} \vert} ^2 {\vert N_{\nu_{k^{\prime}}} \vert} ^2 \left[ \sum\limits_{\alpha=a,b}\frac{1}{\nu_{k}-\epsilon_{\alpha}}\right]\left[\sum\limits_{\alpha=a,b}\frac{1}{\nu_{k^{\prime}}-\epsilon_{\alpha}} \right] \frac{1}{(\nu_{k}-\epsilon_{a})(\nu_{k^{\prime}}-\epsilon_{c})}e^{ i\omega_{k k^{\prime}}t},  \\
F(t)=&\sum\limits_{k} \sum\limits_{k^{\prime}}{\vert N_{\nu_{k}} \vert} ^2 {\vert N_{\nu_{k^{\prime}}} \vert} ^2 \left[ \sum\limits_{\alpha=a,b}\frac{1}{\nu_{k}-\epsilon_{\alpha}}\right]\left[\sum\limits_{\alpha=a,b}\frac{1}{\nu_{k^{\prime}}-\epsilon_{\alpha}} \right] \frac{1}{(\nu_{k}-\epsilon_{b})(\nu_{k^{\prime}}-\epsilon_{c})}e^{ i\omega_{k k^{\prime}}t},
\end{align*}
with the frequency $\omega_{k k^{\prime}}=1/(\nu_{k}-\epsilon_{c})-1/(\nu_{k^{\prime}}-\epsilon_{c})$.
\end{widetext}

By tracing out the third qubit, the reduced density matrices  of three qubit pairs $\rho_{ab},\rho_{ac},\rho_{bc}$ read 
\begin{eqnarray}
\rho_{ab}&=&\frac{C(t)}{2}\vert \uparrow \uparrow \rangle \langle \uparrow \uparrow \vert+\frac{B(t)}{2} \vert \uparrow \downarrow \rangle \langle \uparrow \downarrow \vert+\frac{A(t)}{2} \vert \downarrow \uparrow \rangle \langle \downarrow \uparrow \vert \nonumber \\
&&+\frac{D^{*}(t)}{2} \vert \uparrow \downarrow \rangle \langle \downarrow \uparrow \vert +\frac{D(t)}{2} \vert\downarrow \uparrow  \rangle \langle \uparrow \downarrow \vert,  \nonumber \\
\rho_{ac}&=&\frac{B(t)}{2}\vert \uparrow \uparrow \rangle \langle \uparrow \uparrow \vert+\frac{C(t)}{2} \vert \uparrow \downarrow \rangle \langle \uparrow \downarrow \vert+\frac{A(t)}{2} \vert \downarrow \uparrow \rangle \langle \downarrow \uparrow \vert  \nonumber  \\
&& +\frac{E^{*}(t)}{2} \vert \uparrow \downarrow \rangle \langle \downarrow \uparrow \vert +\frac{E(t)}{2} \vert\downarrow \uparrow  \rangle \langle \uparrow \downarrow \vert, \nonumber  \\
\rho_{bc}&=&\frac{A(t)}{2}\vert \uparrow \uparrow \rangle \langle \uparrow \uparrow \vert+\frac{C(t)}{2} \vert \uparrow \downarrow \rangle \langle \uparrow \downarrow \vert+\frac{B(t)}{2} \vert \downarrow \uparrow \rangle \langle \downarrow \uparrow \vert  \nonumber \\
&&+\frac{F^{*}(t)}{2} \vert \uparrow \downarrow \rangle \langle \downarrow \uparrow \vert +\frac{F(t)}{2} \vert\downarrow \uparrow  \rangle \langle \uparrow \downarrow \vert.  \nonumber 
\end{eqnarray}
Moreover, it's easy to diagonalize  the above three matrices to get their eigenvalues.
The  reduced density matrices of single qubit $\rho_{a},\rho_{b},\rho_{c}$ are given by 
\begin{eqnarray*}
\rho_{a}= &(B(t)/2+C(t)/2)\vert \uparrow \rangle \langle \uparrow \vert+A(t)/2 \vert \downarrow \rangle \langle \downarrow \vert,\\
\rho_{b}= &(A(t)/2+C(t)/2)\vert \uparrow \rangle \langle \uparrow \vert+B(t)/2 \vert \downarrow \rangle \langle \downarrow \vert,\\
\rho_{c}= &(A(t)/2+B(t)/2)\vert \uparrow \rangle \langle \uparrow \vert+C(t)/2 \vert \downarrow \rangle \langle \downarrow \vert.
\end{eqnarray*}

Using the definition of entanglement measures (\ref{Dfsm}) and (\ref{DfC}), we obtain the  quantum mutual information and concurrence, for example the qubit pair $ab$
\begin{align}
S(a:b)&=\gamma_{1}\log_{2}\gamma_{1}+\gamma_{2}\log_{2}\gamma_{2}+\gamma_{3}\log_{2}\gamma_{3}-  A\log_{2}A \nonumber \\
&-B\log_{2}B-(\frac{A}{2}+\frac{C}{2})\log_{2}(\frac{A}{2}+\frac{C}{2})  \nonumber \\
&-(\frac{B}{2}+\frac{C}{2})\log_{2}(\frac{B}{2}+\frac{C}{2}), \\
\mathcal{C}_{ab}&=\max(0,{\lambda_{1}}-{\lambda_{2}}).
\end{align}
Above parameters $\{ \gamma_{i} \}$ and $\{ \lambda_{i}\}$ are respectively the eigenvalues of the density matrix $\rho_{ab}$ and square roots of the matrix $\rho_{ab} \tilde{\rho}_{ab}$ in decreasing order
\begin{align*}
& \gamma_{1,2}=\frac{A+B}{4} \pm  \frac{1}{4}\sqrt{(A-B)^2+\vert D\vert^2}], \\
& \gamma_{3}=\frac{C}{2}, \\
&\lambda_{1,2}^{2}=\frac{1}{4}[(AB+\vert D\vert^2) \pm \sqrt{4AB\vert D\vert^2}].
\end{align*}

 \begin{figure*}[t]
\begin{center}
\includegraphics[scale=0.33]{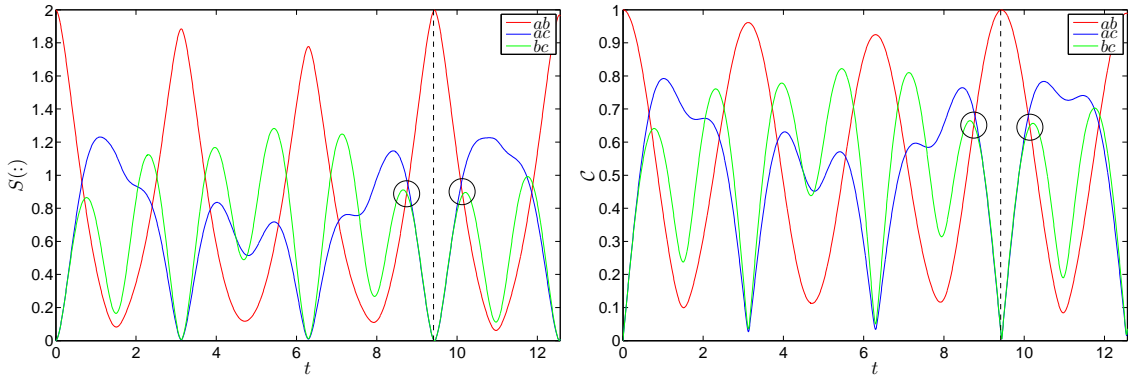}
\end{center} 
\caption{The entanglement dynamics of three qubits with different pairwise interactions. Left panel: Quantum mutual informations  of three qubit pairs evolve in  time. Right panel: Concurrences of  three qubit pairs evolve in time. The red line is the entanglement dynamics of the qubit pair $ab$, the blue line stands for  the entanglement dynamics of the qubit pair $ac$, whereas the green line denotes  the entanglement dynamics of the qubit pair $bc$. Irregular oscillation of the entanglement  dynamics transfers was  observed in both the mutual information and concurrence. The black dashed lines show a nearly perfect revival time of the initial state.}
\label{fig2:dynamics}
\end{figure*}

 \begin{figure}[t]
\begin{center}
\includegraphics[scale=0.35]{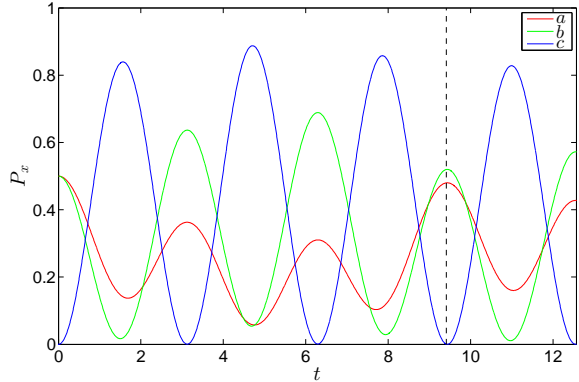}
\end{center} 
\caption{The probability of state $\vert x \rangle$ evolves in time for the  inhomogeneous Hamiltonian (\ref{Ham}). The different color lines  show  the probabilities  of different states $\vert a, b, c \rangle$, Here the  coupling constant $A_{a}=0.5, A_{b}=0.8$. The back dashed line shows a nearly perfect revival time of the initial state. }
\label{fig3:probability}
\end{figure}

From  Fig.\ref{fig2:dynamics}, we observe that  the time evolutions of quantum mutual information and concurrence show a coherence transfer behaviour. 
Initially starting from the entangled state of the qubit pair $ab$,  such a dynamics transfer displays asymmetric feature, i.e. 
the entanglement entropy and concurrence of the  qubit pairs $ac$ and $bc$ oscillate with different frequencies and different magnitudes.
Due to the presence of the  inhomogeneous pairwise  interactions, there does not exist triple intersection point in time evolution of the entanglement dynamics. 
There are the  regions where the entanglements of three qubit pairs are nearly same, see the marked black circles  in Fig.~\ref{fig2:dynamics}.

The probability of state $\vert x \rangle$ with  $x\in \{a,b,c \}$, is defined as 
\begin{equation}
P_{x}=\Tr[\rho_{s}\vert x \rangle \langle x \vert],
\end{equation}
that  measures the probability of projecting the state  $\psi(t)$ onto the  state  $\vert x \rangle$.
In Fig.~\ref{fig3:probability}, we show the probabilities  of the states $\vert a\rangle$,$\vert b \rangle$,$\vert c \rangle$. 
They oscillate anharmonically. A nearly perfect revival of the probabilities of the three states shows the time  $t \approx 9.4$ (dashed line) which is exactly the same as the nearly revival time of  the entropy dynamics, see  Fig.~\ref{fig2:dynamics}.
 This means that at this time the qubit $c$ gets disentangled from the other  two  qubits. 
 However, the system does not completely return  back initial state, since the probability of states $\vert a \rangle$ and $\vert b\rangle$ are  nearly equal. 
  The result originates from both the monogamy of the entanglement of qubit pair $ab$ and the asymmetric pairwise  interactions.

\begin{figure*}[t]
\begin{center}
\includegraphics[scale=0.33]{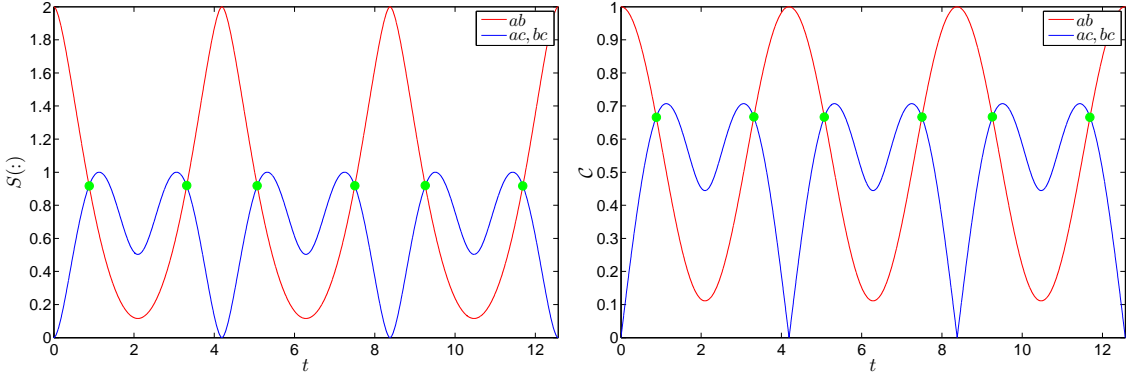}
\end{center} 
\caption{Entanglement dynamics of three qubits for the homogenous Hamiltonian (\ref{Ham2}). Left panel: Quantum mutual information of three qubit pairs $ab,ac,bc$ evolves  time. Right panel: Concurrence of three qubit pairs $ab,ac,bc$ evolves in  time. The  red  lines show  the entanglement of the qubit pair $ab$. The blue lines show  the entanglement of the qubit pairs $ac$ and $bc$. The green dots mark the triple intersection points, i.e. three qubit  pairs have the   equal  mutual information and concurrence. }
\label{fig4:homo}
\end{figure*}

{\em Homogeneous case.} We now  consider the homogenous three interacting qubits with the Hamiltonian 
\begin{equation}
H=
J[\mathbf{s_{a}} \mathbf{s_{c}}+\mathbf{s_{b}} \mathbf{ s_{c}}]. \label{Ham2}
\end{equation}
Here we set the  coupling constant $J=1$ for  a dimensionless unit. 
The Hamiltonian (\ref{Ham2}) can be regarded as  a three-qubit-Heisenberg chain\cite{XiaoGwang}, whose dynamics can be obtained  by the  integrable model \cite{natan2014}.  
Now we may  calculate the  wave function by using a recurrence relation, namely, 
\begin{align*}
&\vert \psi(t) \rangle=\frac{1}{\sqrt{2}}e^{-iHt} (\vert a \rangle +\vert b \rangle ) \\
& =\frac{1}{\sqrt{2}}[\sum_{n=0}^{\infty}\frac{(-it)^n}{n!}H^{n}(\vert a \rangle +\vert b \rangle )] \\
&=\frac{1}{\sqrt{2}}[\sum_{n=1}^{\infty}\frac{(-it)^n}{n!}H^{n-1} \vert c \rangle +(\vert a \rangle +\vert b \rangle )].
\end{align*}
By acting the Hamiltonian  (\ref{Ham2})  on state $\vert c \rangle$ continuously,
we further find  a useful structure for getting  the spectrum of the model
\begin{equation*}
H^{n} \vert c \rangle =\vert c_{1} \rangle \alpha_{1}^{n}+\vert c_{2} \rangle \alpha_{2}^{n},
\end{equation*}
with  $\alpha_{1}=-1,\alpha_{2}=\frac{1}{2}$.
In the above equation the  two states are defined by $\vert c_{1} \rangle =\frac{2}{3}\vert c \rangle -\frac{1}{3}\vert a \rangle -\frac{1}{3}\vert b \rangle$ and $\vert c_{2} \rangle =\frac{1}{3}\vert c \rangle +\frac{1}{3}\vert a \rangle +\frac{1}{3}\vert b \rangle$, where the three basises $\vert a \rangle=\left( \begin{array}{clr}1&0&0 \end{array}\right)^{t} $, $\vert b \rangle=\left( \begin{array}{clr}0&1&0 \end{array}\right)^{t} $, $\vert c \rangle=\left( \begin{array}{clr}0&0&1 \end{array}\right)^{t} $.
Thus the wave function at arbitrary time is given by 
\begin{equation}
\vert \psi(t) \rangle=\frac{1}{\sqrt{2}} \left[ \frac{e^{-i\alpha_{1}t}}{\alpha_{1}} \vert c_{1} \rangle +\frac{e^{-i\alpha_{2}t}}{\alpha_{2}} \vert c_{2} \rangle \right].
\end{equation}
From this  wave function, the density matrix of system is obtained directly 
\begin{equation}
 \rho_{s}= 
\left( \begin{array}{clr}
A (t)& A (t)& B(t) \\
A (t)& A(t) & B(t) \\
B^{*}(t) & B^{*} (t) & C (t)
\end{array}  \right).
\end{equation}
 Three matrix elements $A,B,C$ are respectively given
\begin{align*}
A(t) &=\frac{1}{2} \left[ \frac{1}{9 \alpha_{2}^{2}}+\frac{1}{9 \alpha_{1}^{2}}-\frac{2 \cos[(\alpha_{2}-\alpha_{1})t]}{9 \alpha_{1}\alpha_{2}}\right], \\
B (t)&=\frac{1}{2} \left[ \frac{1}{9 \alpha_{2}^{2}}-\frac{2}{9 \alpha_{1}^{2}}+\frac{2e^{\mathrm{i} (\alpha_{1}-\alpha_{2})t}}{9 \alpha_{1}\alpha_{2}}-\frac{e^{\mathrm{i}(\alpha_{2}-\alpha_{1})t}}{9 \alpha_{1}\alpha_{2}}\right], \\
C (t)&=\frac{1}{2} \left[ \frac{1}{9 \alpha_{2}^{2}}+\frac{4}{9 \alpha_{1}^{2}}+\frac{4 \cos[(\alpha_{2}-\alpha_{1})t]}{9 \alpha_{1}\alpha_{2}}\right]. 
\end{align*}
We further obtain the reduced density matrices of three qubit pairs $\rho_{ab},\rho_{ac},\rho_{bc}$  by tracing out the third qubit
\begin{eqnarray*}
\rho_{ab}= &C(t)\vert \uparrow \uparrow \rangle \langle \uparrow \uparrow \vert+A(t) \vert \uparrow \downarrow \rangle \langle \uparrow \downarrow \vert+A(t) \vert \downarrow \uparrow \rangle \langle \downarrow \uparrow \vert  \\
&+A(t) \vert \uparrow \downarrow \rangle \langle \downarrow \uparrow \vert+A(t) \vert\downarrow \uparrow  \rangle \langle \uparrow \downarrow \vert, \\
\rho_{ac}= &A(t)\vert \uparrow \uparrow \rangle \langle \uparrow \uparrow \vert+C(t) \vert \uparrow \downarrow \rangle \langle \uparrow \downarrow \vert+A(t) \vert \downarrow \uparrow \rangle \langle \downarrow \uparrow \vert \\
&+B^{*}(t) \vert \uparrow \downarrow \rangle \langle \downarrow \uparrow \vert+B(t) \vert\downarrow \uparrow  \rangle \langle \uparrow \downarrow \vert, \\
\rho_{bc}= &A(t)\vert \uparrow \uparrow \rangle \langle \uparrow \uparrow \vert+C(t) \vert \uparrow \downarrow \rangle \langle \uparrow \downarrow \vert+A(t) \vert \downarrow \uparrow \rangle \langle \downarrow \uparrow \vert \\
&+B^{*}(t) \vert \uparrow \downarrow \rangle \langle \downarrow \uparrow \vert+B(t) \vert\downarrow \uparrow  \rangle \langle \uparrow \downarrow \vert.
\end{eqnarray*}
Note that here we used the same notations for these functions  $A(t),\,B(t),\,C(t)$  as being used in the  inhomogeneous case.

It's easy to diagonalize the above three matrices to obtain their eigenvalues. 
Moreover, the reduced density matrices of the single qubit $\rho_{a},\, \rho_{b},\, \rho_{c}$ are given by 
\begin{eqnarray*}
\rho_{a}= &(A(t)+C(t))\vert \uparrow \rangle \langle \uparrow \vert+A(t) \vert \downarrow \rangle \langle \downarrow \vert, \\
\rho_{b}= &(A(t)+C(t))\vert \uparrow \rangle \langle \uparrow \vert+A(t) \vert \downarrow \rangle \langle \downarrow \vert, \\
\rho_{c}= &2A(t)\vert \uparrow \rangle \langle \uparrow \vert+C(t) \vert \downarrow \rangle \langle \downarrow \vert. \\ 
\end{eqnarray*}
There are only diagonal elements in the reduced density matrix of single qubits $\rho_{a},\rho_{b},\rho_{c}$ due to the conserved magnetization. 
Accoding to the definition of entanglement measure, for instance,  the quantum mutual information and concurrence  of qubit pair $ab$
\begin{align}
S(a:b)&=2A+C\log_{2}C-2(A+C)\log_{2}(A+C),  \\
\mathcal{C}_{ab}&=\max(0,2A).
\end{align}
We also can derive the probabilities  $P_{a,b,c}$ of the three states  like what discussed  in the inhomogeneous case.

 \begin{figure}[t]
\begin{center}
\includegraphics[scale=0.35]{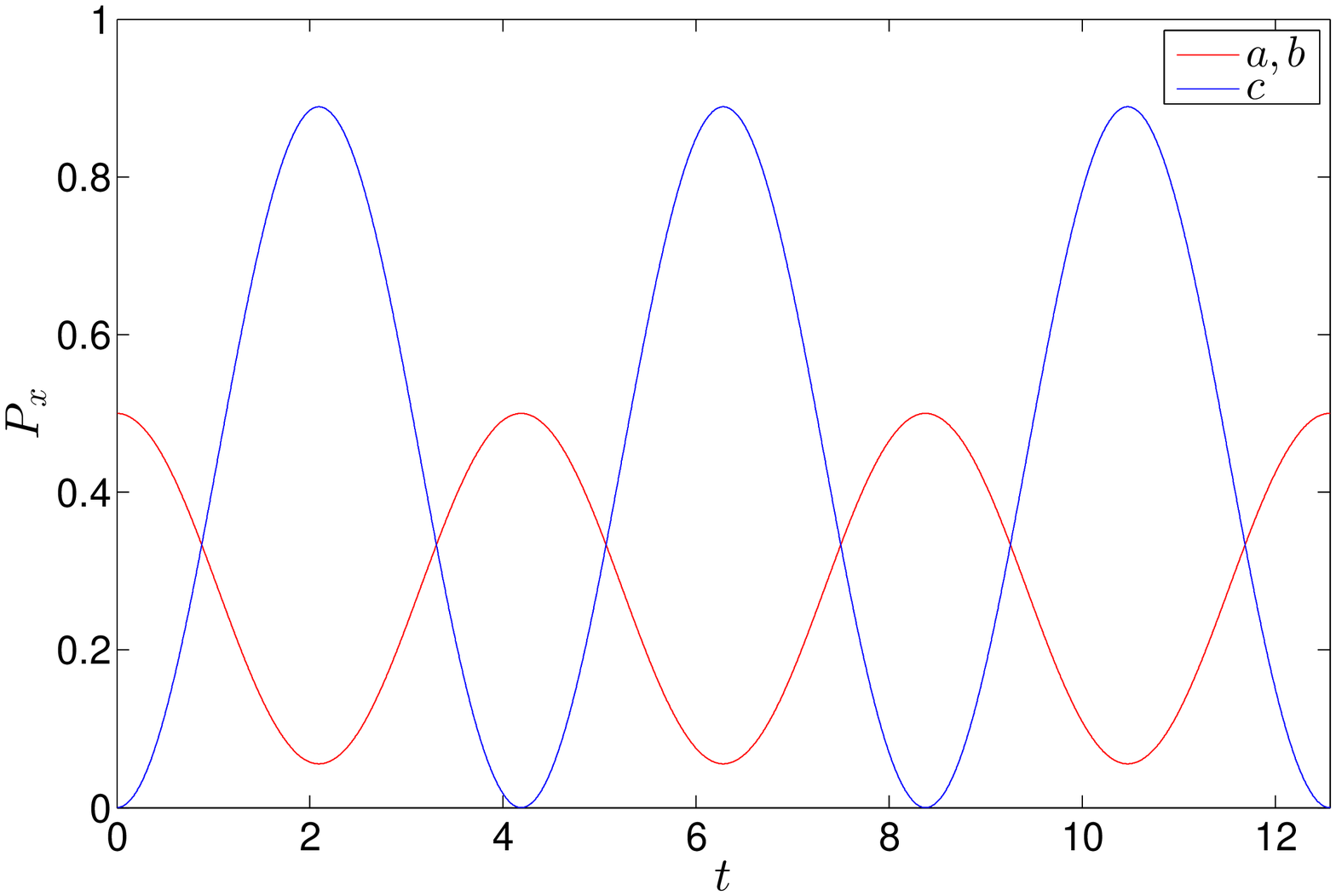}
\end{center} 
\caption{The probability of state $\vert x \rangle$  evolve with time for homogeneous case. The red line is the probability of state $\vert a \rangle$ or  $\vert b \rangle$, blue line is the probability of state $\vert c \rangle$}
\label{fig5:probability}
\end{figure}

Fig.\ref{fig4:homo} shows the entanglement entropy and concurrence of the homogenous Hamiltonian (\ref{Ham2}).  
We observe that the genuine three-qubit  entangled  state is naturally induced through two pairwise interactions $ac$ and $bc$. 
There exist some  special states at  which the entanglements of three qubit pairs are the same, see the marked green dots in Fig.~\ref{fig4:homo}.
The times when  the three pairwise  entangled  states are  equal  satisfy  the relation $t_{e}=\pm \frac{2}{3}\arccos(\frac{1}{4})+\frac{4}{3}n \pi$, here $n \in \{0,1,2,\cdots \}$.
While at these points  the probabilities of the single states $\vert a\rangle$,$\vert b \rangle$,$\vert c \rangle$ are also the same, see  Fig.~\ref{fig5:probability}.
In contrast to the inhomogenous case  Fig.~\ref{fig3:probability},  the probability $P_{a},P_{b},P_{c}$ oscillate periodically due to the homogeneous pairwise interaction.
The state consisting of  three equally  entangled  states is called the $W$ state, where the three qubits are a equally weighted  superposition and the norm of the off-diagonal element $B$ in density matrix  is $1/3$. 
 We can prove that the quantum mutual information of  the  equally entangled state is same as the quantum mutual information of the $W$ state  with the entanglement entropy  $S_{w}=\log_{2}(3)-\frac{2}{3}\approx 0.9183$ for two qubits pair.
It is also  easy to check the concurrence $\mathcal{C}=\frac{2}{3}$ for both. 
This is a very interesting feature that dynamical evolution of the  pairwise entangled state can produce  a three-partite entangled  $W$ state.
In contrast to the smooth time evolution of the quantum mutual information, the concurrence shows  sharp changes  at certain times, see the blue line in the right panel of Fig.~\ref{fig4:homo}.
The evolutions of quantum mutual information and concurrence  reveal a very interesting  features of quantum entanglement transfer. 

\emph{Experimental Proposal}
Finally, we propose an experimental scheme to test the above entanglement dynamics  by using Rydberg atom \cite{M. Saffman}. 
One can  use three $^{87}$Rb atoms to simulate the entanglement dynamics transfer in the homogeneous Hamiltonian (\ref{Ham2})  and  use two $^{87} $Rb atoms and one $^{23} Na$ atom or three heteronuclear Rb atoms to simulate the such a dynamics transfer in the  inhomogeneous Hamiltonian (\ref{Ham}). 
For the  homogeneous case, we estimate the time of the first equally  entangled state point, $t_{1}=\frac{2}{3J}\arccos(\frac{1}{4}) \approx 0.8787/J$. 
Here we used the data of spin exchange coupling constant  given in  \cite{Barredo:2015}. 
The spin exchange coupling reads $J=C_{3}^{\exp}/R^3$ with the parameter setting $C_{3}^{\exp}=7950 \pm 130 \MHz {\mu m}^3, R=30 \mu m$. 
Thus the estimated time of the first equally  entangled state is at  $t_{1}\approx 2.9843 \mu s$, which can be  accessible experimentally.

{\em In summary.} We have studied  the quantum  entanglement dynamics transfer in  three pairwise interacting qubits. 
We have analytically calculated time evolutions of wave function,  density matrix and entanglement entropy for the system. 
We have found   that pairwise interactions may induce  a genuine three-qubit entangled  state during time evolution. 
In such a three-qubit entangled state, the mutual entanglement entropies can be  equally weighted depending on the choices of the pairwise interactions.
 The evolution of quantum mutual information is  a smooth function of time. 
 The concurrence displays  some sharp changes at some points. 
The entanglement dynamics transfer in the inhomogeneous system  is anharmonic. 
In this case, the  initial state can not completely return  back  even the entanglement of the initial qubit pair $ab$  reaches the  maximum.

\noindent

{\em Acknowledgments.} This work is supported by the key NSFC grant No.\ 11534014 and the National Key R\&D Program of China  No. 2017YFA0304500.


\begin{thebibliography}{99}
\bibitem{Barredo:2015} Barredo D, Labuhn H,  Ravets S, Lahaye T , Browaeys A, and Adams C S  2015 {Phys. Rev. Lett} {\bf 114} 113002 
\bibitem{Rosenfeld:2018} Rosenfeld E L , Pham L M , Lukin M D and Walsworth  R L  2018 {Phys. Rev. Lett} {\bf 120}  243604 

\bibitem{APolkRMP}  Polkovnikov A , Sengupta K, Silva A , and Vengalattore M  2011 {Rev. Mod. Phys.} {\bf 83}  863
\bibitem{Fazio} Amico L , Fazio R, Osterloh A , and Vedral V  2008 {Rev. Mod. Phys.} {\bf 80} 517
\bibitem{Qmetro} Giovannetti V , Lloyd S , and Maccone L 2006 {Phys. Rev. Lett.} {\bf 96}  010401 

\bibitem{DJwang} Li X K , Zhu B , He X D , Wang F D, Guo M Y, Xu Z F , Zhang S Z , and Wang D J   2015 {Phys. Rev. Lett.} {\bf 114}  255301
\bibitem{MSZhan} Zeng Y , Xu P , He X D , Liu Y Y, Liu M , Wang J, Papoular D  J, Shlyapnikov G V , and Zhan M S  2017 {Phys. Rev. Lett.} {\bf 119} 160502 
\bibitem{nature} DiCarlo L, Reed M D, Sun L,  Johnson B R , Chow J M , Gambetta J M , Frunzio L , Girvin S M , Devoret M H and Schoelkopf R J  2010 {Nature} {\bf 467} 574–578
\bibitem{GcGuo} Wang B C , Cao G , Li H O , Xiao M , Guo G C , Hu X D , Jiang H W , and Guo G P   2017 {Phys. Rev. Applied.} {\bf 8}  064035
\bibitem{Nielsen} Nielsen M  A  and Chuang I  L , \emph{Quantum Computation and Quantum Information} (Cambridge University Press, Cambridge, England, 2000).
\bibitem{Wootters97} Hill S  and Wootters W K  1997 {Phys. Rev. Lett.} {\bf 78} 5022
\bibitem{Wootters98}  Wootters W K  1998 {Phys. Rev. Lett.} {\bf 80} 2245 
\bibitem{Wootters00} Coffman V, Kundu J and Wootters W K  2000  {Phys. Rev. A.} {\bf 61}  052306 

\bibitem{J.Dukelsky2004} Dukelsky J, Pittel S, Sierra G  2004 {Rev. Mod. Phys.} {\bf 76}, 643 
\bibitem{H.-Q. Zhou}Zhou H Q, Links J , McKenzie R H , and Gould M D  2002 {Phys. Rev. B.} {\bf 65}  060502(R) 
\bibitem{Dloss02} Khaetskii A V, Loss D, Glazman  L  2002 {Phys. Rev. Lett.} {\bf 88}  18
\bibitem{Dloss03} Khaetskii A V, Loss D,  Glazman  L  2003 {Phys. Rev. B.} {\bf 67}  195329
\bibitem{Bortz07} Bortz M, Stolze J  2007 {Phys. Rev. B.} {\bf 76}  014304
\bibitem{AFprb} Araby O E , Gritsev V , and Faribault A   2012  {Phys. Rev. B.} {\bf 85} 115130

\bibitem{XiaoGwang} Wang X G  2001 {Phys. Rev. A.} {\bf 64} 012313 
\bibitem{natan2014} Liu W S ,  Andrei N  2014 {Phys. Rev. Lett.} {\bf 112} 257204
\bibitem{M. Saffman} Saffman M, Walker T  G, and  M$o\!\!\!\slash$lmer K  2010 {Rev. Mod. Phys.} {\bf 82} 2313
\end{thebibliography}
\end{document}